\documentclass{myws-ijqi2}
\usepackage{verbatim,amssymb} 

\newcommand{\mysubsectionapp}[1]{\addtocounter{subsection}{1}\subsection*{\Alph{section}\arabic{subsection}. #1}} 

\newcommand{\enproof}{\hfill $\Box$ \vspace*{1ex}}
\newcommand{\enlem}{\end{lemma}} 
\newcommand{\closedef}{\end{definition}} 
\newcommand{\enth}{\end{theorem}} 
\newcommand{\enprop}{\end{proposition}} 
\newcommand{\encond}{\end{condition}} 
\newcommand{\exam}[1]{\begin{example}\label{ex:#1}}
\newcommand{\enexam}{\QED\end{example}}
\newcommand{\beremark}[1]{\begin{remark}\label{rmk:#1}}

\newcommand{\mymathbb}[1]{{\mathbb #1}} 
\newcommand{\mymathsf}[1]{{\mathsf{#1}}} 

\newcommand{\dmn}{d}

\newcommand{\cA}{{\cal A}}

\newcommand{\cB}{{\cal B}}

\newcommand{\cC}{{\cal C}}

\newcommand{\cD}{{\cal D}}
\newcommand{\cE}{{\cal E}}

\newcommand{\myF}{{\mymathbb{F}_{\dmn}}} 
\newcommand{\cG}{{\cal G}}

\newcommand{\cN}{{\cal N}}

\newcommand{\sH}{\mymathsf{H}} 
\newcommand{\cI}{{\cal I}}

\newcommand{\cL}{{\cal L}}
\newcommand{\sL}{\mymathsf{L}}
\newcommand{\cM}{{\cal M}}
\newcommand{\sM}{\mymathsf{M}}

\newcommand{\sQ}{\mymathsf{Q}}

\newcommand{\cR}{{\cal R}}
\newcommand{\cS}{{\cal S}}

\newcommand{\cX}{{\cal X}}

\renewcommand{\phi}{\varphi} 
\renewcommand{\subset}{\subseteq}
\renewcommand{\tilde}{\widetilde}
\renewcommand{\hat}{\widehat}
\newcommand{\Bar}{\overline}  
\renewcommand{\bar}{\Bar}

\newcommand{\dss}{\displaystyle}
\newcommand{\mbm}[1]{\mbox{\boldmath $#1$}}

\newcommand{\SINT}{\mymathbb{Z}}

\newcommand{\Capa}{\mymathsf{Q}} 

\newcommand{\Expe}{\mymathbb{E}}
\newcommand{\Dopt}{\mymathsf{D}} 

\newcommand{\transp}{^{\rm T}}

\newcommand{\tnsr}{\otimes}

\newcommand{\lag}{\langle}
\newcommand{\rag}{\rangle}
\newcommand{\crd}[1]{|#1|}
\newcommand{\bra}[1]{\lag #1 |}
\newcommand{\ket}[1]{| #1 \rag}

\newcommand{\syp}[2]{( #1,  #2 )_{\rm sp}}
\newcommand{\perpsyp}{\perp}

\newcommand{\Hch}{{\sH}}
\newcommand{\Hgn}{{\sH}}
\newcommand{\Hgen}{{\sH^{\tnsr n}}}
\newcommand{\Hgem}{{\sH^{\tnsr \genn}}}

\newcommand{\Bop}{\sL} 
\newcommand{\Hcd}{\cC} 

\newcommand{\Cq}{\Hcd}
\newcommand{\Cqarg}[1]{\Hcd^{#1}} 

\newcommand{\Fen}{F_{\rm e}}

\newcommand{\imu}{\sqrt{-1}} 
\newcommand{\Ebe}{N}

\newcommand{\ketbe}[1]{\ket{#1}}
\newcommand{\phasebe}{\omega}
\newcommand{\Xbe}{X}
\newcommand{\Zbe}{Z}

\newcommand{\Jcr}{J}  
\newcommand{\Icr}{\Jcr}

\newcommand{\Cso}{L}

\newcommand{\vara}{l} 

\newcommand{\varu}{u} 

\newcommand{\varb}{\vara} 

\newcommand{\vars}{s} 
\newcommand{\varss}{a}
\newcommand{\varsss}{a'}
\newcommand{\vartt}{b}
\newcommand{\varttt}{b'}

\newcommand{\varnin}{n} 
\newcommand{\varN}{\nu}   

\newcommand{\varm}{\exgnn} 
\newcommand{\varnn}{n}
\newcommand{\genn}{n} 
\newcommand{\exgnn}{n} 
\newcommand{\exgenm}{m} 
\newcommand{\genk}{k} 

\newcommand{\spn}{\mymathsf{span}\,}
\newcommand{\Proj}{\Pi} 

\newcommand{\mms}[1]{\pi_{#1}} 
\newcommand{\ghb}[1]{\Bar{#1}}
\newcommand{\Id}{\cI} 

\newcommand{\varsp}{s} 
\newcommand{\varrc}{x} 

\newcommand{\varM}{l} 
\newcommand{\varDK}{K} 
\newcommand{\varNK}{M}

\newcommand{\Dec}{\cR} 
\newcommand{\DKO}{K} 

\newcommand{\hMsub}[2]{\sM_{#1}({#2})} 
\newcommand{\hMsubbar}[2]{\bar{\sM}_{#1}({#2})} 
\newcommand{\hMsubinv}[2]{\sM^{-1}_{#1}({#2})} 
\newcommand{\shs}{\rho} 

\newcommand{\dbar}[1]{\bar{\bar{#1}}}

\newcommand{\myFpower}[1]{\mymathbb{F}_{\dmn}^{#1}}

\newcommand{\Zdpower}[2]{\mymathbb{F}_{#1}^{#2}} 

\newcommand{\Qpl}[1]{\sQ^+} 
\newcommand{\Qmi}[1]{\sQ^-}

\newcommand{\Eve}{\cA}
\newcommand{\Bch}{\cA} 
\newcommand{\Bchgen}{\cA} 
\newcommand{\Enr}{\cB} 

\newcommand{\Jsp}{J}
\newcommand{\twl}[1]{\tilde{#1}}
\newcommand{\twA}{\twl{\Bchgen}} 

\newcommand{\codesubs}[1]{\cC^{(#1)}} 
\newcommand{\codesubsbar}[1]{\cC^{'(#1)}} 
\newcommand{\gfL}{g} 
\newcommand{\Hchan}{\Hch_{\rm c}}

\newcommand{\Hout}{\Hch_{\rm o}}
\newcommand{\Hchann}{\Hch^{\tnsr n}} 

\newcommand{\Hbg}{C} 
\newcommand{\Ebg}{\cE}
\newcommand{\Dbg}{\cD}
\newcommand{\Rin}{r} 
\newcommand{\Pbg}{P_{\Enr}}

\newcommand{\QCp}{\Capa_{\rm p}}
\newcommand{\QCpd}{\Capa_{{\rm p},\dmn}}
\newcommand{\QCe}{\Capa_{\rm e}}
\newcommand{\QCed}{\Capa_{{\rm e},\dmn}}
\newcommand{\rpUnoarg}{U}
\newcommand{\rpU}[1]{\rpUnoarg_{#1}}
\newcommand{\fs}[2]{\eta(#1,#2)}
\newcommand{\anotherfs}[2]{\zeta(#1,#2)}
\newcommand{\newx}{z}
\newcommand{\newz}{w}
\newcommand{\newX}[1]{\bar{X}^{#1}} 
\newcommand{\newZ}[1]{\bar{Z}^{#1}} 
\newcommand{\oldx}{a}
\newcommand{\oldz}{b}
\newcommand{\Ebeh}[1]{\newX{#1}}
\newcommand{\Ebeg}[1]{\newZ{#1}}
\newcommand{\Phs}{\Phi_s}

\newcommand{\mycite}[1]{Ref.~\refcite{#1}}

\newcommand{\myspcite}[1]{\cite{#1}} 

\begin{document}

\markboth{Mitsuru Hamada}
{Notes on the Fidelity of Symplectic Quantum Error-Correcting Codes}

%
\catchline{}{}{}{}{}
%

\title{NOTES ON THE FIDELITY OF SYMPLECTIC QUANTUM ERROR-CORRECTING CODES}

\author{MITSURU HAMADA
}

\address{Quantum Computation and Information Project, ERATO Program,\\
     Japan Science and Technology Agency,\\
201 Daini Hongo White Bldg.,
      5-28-3, Hongo, Bunkyo-ku, Tokyo 113-0033, Japan}

\maketitle

\begin{abstract}
Two observations are given on the fidelity of 
schemes for quantum information processing.
In the first one, we 
show that the fidelity of a symplectic (stabilizer) code,
if properly defined, exactly equals 
the `probability' of the correctable errors for general quantum
channels. The second observation states that for any coding rate 
below the quantum capacity, 
exponential convergence of the fidelity of some codes
to unity is possible.
\end{abstract}


\section{Introduction}

Two observations are given in this paper on the fidelity of 
schemes for quantum information processing, especially on that of 
quantum codes and entanglement distillation protocols.
In the first one, we give a formula for the fidelity of symplectic (stabilizer) codes\cite{crss97,crss98,gottesman96}.
While relating the fidelity of symplectic codes
with the `probability' of correctable errors
for channels represented by trace-preserving
completely positive (TPCP) maps was already done in the
literature\myspcite{KnillLaflamme97,GottesmanPreskill01,hamada01g},
this work shows that the fidelity, if properly defined, exactly equals 
the `probability' of the correctable errors for general quantum
channels. This formula is also useful for assessing the security of quantum key distribution (QKD) protocols\cite{hamada03s}.
In fact,
one of the motivations for analyzing the fidelity of symplectic codes was
to prove the security 
of the Bennett-Brassard 1984 (BB84) QKD protocol\myspcite{BennettBrassard84}
or its analogs along the lines of Shor and Preskill\myspcite{ShorPreskill00,GottesmanPreskill01,hamada03s}.

The second observation is related to the problem of the quantum capacity of noisy quantum channels\myspcite{Shor02msri,devetak03}.
It states that for any coding rate below the quantum
capacity, exponential convergence of the fidelity of some codes 
to unity is possible.

This paper is organized as follows.
In Section~\ref{ss:wb}, several basic notions such as Weyl's unitary basis 
are introduced. Section~\ref{ss:sc} contains the formula for
the fidelity of symplectic codes, which is applied to entanglement distillation
in Section~\ref{ss:distill}. 
Sections~\ref{ss:exp} and \ref{ss:rc}, respectively, 
contain the observation on exponential
convergence of fidelity and a known lemma to be used in the subsequent section,
where the observation is proved. 
Sections~\ref{ss:exp_distill} and \ref{ss:conc}
contain a remark and a summary, respectively.
Two appendices are given to explicate the basics of symplectic codes
and to give a technical argument on the capacity, respectively.

\section{Basic Notions \label{ss:wb}}

\subsection{Terminology and Notation}

We will treat copies of a quantum system 
described with $\Hgn$, $\dmn=\dim \Hgn < +\infty$.
A composite system consisting of $n$ such copies is sometimes called an
$n$-quantum-($\dmn$-ary-)digit system. 
The set of all linear maps from a Hilbert space $\Hgn$ into itself is denoted by
$\Bop(\Hgn)$.
Hereafter throughout, it is assumed that
$\Hgn$ is a Hilbert space whose dimension $\dmn$
is a prime number, though the results in this section
are true for any integer $\dmn \ge 2$.
We assume this because the structure of vector spaces
over the finite field $\myF =\SINT/\dmn\SINT$ will be exploited.
For two subsets $A$ and $B$ of an additive group,
$A+B$ denotes $\{a+b \mid a\in A, b\in B\}$,
and $a+B$ denotes $\{a \}+ B$. 

In this paper, the way to specify quantum codes varies according to the 
context. For most parts, a quantum code 
indicates a pair $(\cC,\cR)$ consisting of a
code subspace $\cC$ of $\Hgn^{\tnsr n}$ and a recovery
operator $\cR$;
sometimes $\cC$ alone is called a quantum code.
A more general definition allowing encoding maps
will appear in a later section.

\subsection{The Weyl Basis}

A representation 
$\rpUnoarg: \cG \ni x \mapsto \rpU{x} \in \Bop(\Hgn)$ 
of a group $\cG$ usually indicates
one with the property $\rpU{x+y}=\rpU{x}\rpU{y}$, $x,y\in\cG$.
However, in quantum mechanics, 
vectors in $\Bop(\Hgn)$ proportional to each other
stand for a single quantum state, so that it is natural to weaken
the stipulation $\rpU{x+y}=\rpU{x}\rpU{y}$ to 
that $\rpU{x+y}= \fs{x}{y} \rpU{x}\rpU{y}$, $x,y\in \cG$,
for some collection of complex numbers $\fs{x}{y}$, $x,y\in \cG$.
If $\rpUnoarg$ satisfies the weaker assumption, it is called
a ray (projective) representation. 

Weyl\myspcite{weyl28} introduced
two unitary operators, $X$ and $Z$, on $\Hgn$ satisfying the property
\begin{equation}\label{eq:primCR}
XZ = \omega ZX,
\end{equation}
with $\omega$ being a primitive $\dmn$-th root of unity
to give a unitary ray representation, $\Ebe$, of $\cX=\myFpower{2}$,
the 2-dimensional numerical vector space.
A concrete form of $\Ebe$ can be given as follows.
Fix an orthonormal basis 
$\{ \ketbe{0},\dots, \ketbe{\dmn-1} \}$ of $\Hch$.
Define $X$ and $Z$ by
\begin{equation}\label{eq:error_basis}
\Xbe \ketbe{\varss}  = \ketbe{\varss-1}, \quad
\Zbe \ketbe{\varss} = \phasebe^\varss \ketbe{\varss}, \quad \varss\in \myF.
\end{equation}
We define $\Ebe$ by 
\begin{equation}\label{eq:error_basis_d2}
\Ebe_{(\varss,\vartt)} = \imu^{\varss\vartt} \Xbe^\varss \Zbe^\vartt,\quad (\varss,\vartt)\in\cX
\end{equation}
for $\dmn=2$, and by 
\begin{equation}\label{eq:error_basis_d3}
\Ebe_{(\varss,\vartt)} = \Xbe^\varss \Zbe^\vartt,\quad (\varss,\vartt)\in\cX
\end{equation}
for $\dmn>2$.
For $\dmn=2$, $\Ebe_{(\varss,\vartt)}$, $(\varss,\vartt)\ne(0,0)$, are the Pauli operators.
Note that there are many systems of
complex numbers $\anotherfs{\varss}{\vartt}$ of modulus 1
such that $\anotherfs{\varss}{\vartt}X^{\varss}Z^{\vartt}$ 
is a ray representation of $\cX$.
Using the factor $\imu^{\varss\vartt}$ in the case of $\dmn=2$ 
is for a technical reason (\ref{app:sc}, Section~A.5). 
It is remarked that Weyl actually derived
the concrete representation in (\ref{eq:error_basis_d3}) from 
(\ref{eq:primCR})
with more natural stipulations such as the irreducibility of 
$\Ebe$.

We identify $((x_1,z_1),\dots,(x_{\genn},z_{\genn}))\in \cX^{\genn}$
with $(x_1,z_1,\dots,x_{\genn},z_{\genn})\in\myFpower{2\genn}$.
To cope with composite quantum systems,
we write 
$\Ebe_{y}=\Ebe_{y_1}\tnsr
\cdots \tnsr \Ebe_{y_n}$, where $y =(y_1,\dots,y_n)\in\cX^n$,
and $\Ebe_{\Jcr}=\{\Ebe_y \mid y\in\Jcr \}$, where $\Jcr\subset\myFpower{2n}$.
We call the operators $\Ebe_y$ Weyl unitaries and
the system $\{ \Ebe_{y} \}_{y\in\myFpower{2n}}$ Weyl basis. 
An important property of the Weyl basis
is the commutation relation 
\begin{equation}\label{eq:wcr}
 \Ebe_{y} \Ebe_{y'} = \omega^{\syp{y}{y'}} \Ebe_{y'} \Ebe_{y},
\end{equation}
where
\begin{equation}\label{eq:syp}
\syp{y}{y'} = \sum_{i=1}^{n} x_i z_i' - z_i x_i' 
\end{equation}
for $y=(x_1,z_1,\dots,x_{\genn},z_{\genn})$ and 
$y'=(x'_1,z'_1,\dots,x'_{\genn},z'_{\genn}) \in \Zdpower{\dmn}{2\genn}$.
The commutation relation (\ref{eq:wcr}) follows from 
\begin{equation}\label{eq:prW}
\Ebe_{(\varss,\vartt)} \Ebe_{(\varsss,\varttt)} = \phasebe^{-\vartt\varsss }
\Ebe_{(\varss+\varsss,\vartt+\varttt)}, \quad \varss,\vartt,\varsss,\varttt\in\myF,
\end{equation}
which in turn follows from the primitive relation (\ref{eq:primCR}), and 
the map that sends $(y,y')$ to $\syp{y}{y'}$ in (\ref{eq:syp})
is known as a symplectic bilinear form.
The relation (\ref{eq:wcr}) implies that
{\em $\syp{x}{y}=0$ if and only if $\Ebe_{x}$ and $\Ebe_{y}$ commute.}

We have a lemma~\myspcite{schwinger60,werner01}.
\begin{lemma}\label{lem:Phix}
The vectors
\begin{equation*} 
\ket{\Psi_{y}} = 
\frac{1}{\sqrt{\dmn^n}} \sum_{\vara\in\myFpower{n}} \ket{\vara} \tnsr \Ebe_y \ket{\vara}, \quad y \in\myFpower{2n}
\end{equation*}
where $\ket{(l_1,\dots,l_n)}=\ket{l_1}\tnsr\dots\tnsr\ket{l_n}$, 
form an orthonormal basis of $\Hgen \tnsr \Hgen$.
\end{lemma}
Note that 
putting $\ket{\Psi}=\ket{\Psi_{0^{2n}}}$ for the zero vector $0^{2n}$
in $\myFpower{2n}$, 
we can rewrite $\ket{\Psi_y}$ as $(I \tnsr \Ebe_y) \ket{\Psi}$.
The zero vector $0^{m}\in\myFpower{m}$ will be sometimes abbreviated as $0$
if there is no fear of confusion.

\subsection{Choi's Matrix}

A simple but helpful tool in quantum information theory is
the following one-to-one map of Choi\myspcite{choi75}
between the CP maps on $\Bop(\Hgen)$ and the positive semi-definite
operators in $\Bop(\Hgen\tnsr\Hgen)$:
\begin{equation}\label{eq:choi}
\hMsub{n}{\Bchgen} = [\Id \tnsr \Bchgen](\ket{\Psi} \bra{\Psi}),
\end{equation}
where $\Id$ is the identity map on $\Bop(\Hgen)$.
In fact, Choi introduced $\dmn^{n}\hMsub{n}{\Bchgen}$
in the matrix form (with more flexibility on dimensionality) 
to yield fundamentals of CP maps.

According to Theorem~1 of Choi\myspcite{choi75}, 
if $\shs_{\genn}=\hMsub{\genn}{\Bchgen}$ is written as
\begin{equation}\label{eq:MatrixBellBasis}
\shs_n = \sum_{y,z\in\Zdpower{\dmn}{2n}} \alpha_{y,z} 
\ket{\Psi_{y}} \bra{\Psi_{z}},
\end{equation}
or equivalently as
\[
\shs_{\genn} = \frac{1}{\dmn^{\genn}} \sum_{l,m\in\Zdpower{\dmn}{\genn}}\sum_{y,z\in\Zdpower{\dmn}{2\genn}} 
\ket{l}\bra{m} \tnsr \alpha_{y,z} \Ebe_y \ket{l} \bra{m} \Ebe_z^{\dagger}, 
\]
then the CP map $\Bchgen$ is represented as
\begin{equation}\label{eq:choiN}
\Bchgen: \sigma \mapsto \sum_{y,z\in\Zdpower{\dmn}{2\genn}} 
\alpha_{y,z} \Ebe_y \sigma \Ebe_z^{\dagger}.
\end{equation}
This immediately follows from the fact that Choi's matrix, viz.,
the matrix of $\dmn^{n}\hMsub{n}{\Bchgen}$ with respect to the basis
$\{ \ket{l}\tnsr\ket{m}\}_{l,m}$, is
the $\dmn^{\genn} \times \dmn^{\genn}$
block matrix whose $(l,m)$-entry is the 
$\dmn^{\genn} \times \dmn^{\genn}$ matrix of $\Bchgen(\ket{l}\bra{m})$.

\subsection{Discrete Twirling}

We begin with proving the following formula for discrete
twirling (Appendix~A of \mycite{bennett96m}, \mycite{hamada03t}): 
For an operator $\shs_n \in \Bop(\Hgen)$ in
(\ref{eq:MatrixBellBasis}),
we have
\begin{equation}\label{eq:twirling}
\frac{1}{\dmn^{2n}} \sum_{x\in\Zdpower{\dmn}{2n}}
(\overline{\Ebe_{x}} \tnsr \Ebe_x) \shs_n
(\overline{\Ebe_{x}} \tnsr \Ebe_x)^{\dagger}
=\sum_{y\in\Zdpower{\dmn}{2n}} \alpha_{y,y}\ket{\Psi_y}\bra{\Psi_y}
\end{equation}
where $\bar{U}$ is the complex conjugate of $U$, viz.,
the element $\bra{l}\bar{U}\ket{m}$ is the complex conjugate of
$\bra{l}U\ket{m}$ for $l,m\in\Zdpower{\dmn}{n}$.

{\em Proof of (\ref{eq:twirling})}\/.
Put
\[
\shs_n'=\frac{1}{\dmn^{2n}} \sum_{x\in\myFpower{2n}}
(\overline{\Ebe_{x}}\tnsr \Ebe_x) \shs_n (\overline{\Ebe_{x}}\tnsr \Ebe_x)^{\dagger}.
\]
Then,
\begin{eqnarray}
\shs_n'&=& \frac{1}{\dmn^{2n}} \sum_{x,y,z\in\myFpower{2n}} \alpha_{y,z}
(\overline{\Ebe_{x}}\tnsr \Ebe_x\Ebe_y) \ket{\Psi}\bra{\Psi} (\overline{\Ebe_{x}}\tnsr \Ebe_x\Ebe_z)^{\dagger} \nonumber\\
&=& \frac{1}{\dmn^{2n}} \sum_{x,y,z\in\myFpower{2n}} \alpha_{y,z}(I\tnsr \Ebe_x\Ebe_y\Ebe_x^{\dagger}) 
\ket{\Psi}\bra{\Psi} (I \tnsr \Ebe_x\Ebe_z\Ebe_x^{\dagger})^{\dagger}, \label{eq:twirled}
\end{eqnarray}
where we used the relation
\begin{equation}\label{eq:me_t}
(A \tnsr I)\ket{\Psi}=(I\tnsr A\transp)\ket{\Psi}
\end{equation}
with $A\transp$ 
being the transpose of $A$ with respect to $\{ \ket{j} \}$, which means that
if $A=\sum_{l,m} a_{l,m}\ket{l}\bra{m}$, then
$A\transp=\sum_{l,m} a_{m,l}\ket{l}\bra{m}$.
Using (\ref{eq:wcr}), we then have
\[ 
\shs_{\varnn}'=
\frac{1}{\dmn^{2n}} \sum_{x,y,z\in\myFpower{2n}} \alpha_{y,z} 
\omega^{\syp{x}{y-z}}
(I\tnsr \Ebe_y) \ket{\Psi}\bra{\Psi} (I \tnsr \Ebe_z)^{\dagger}.
\] 
Since
\[ 
\sum_{x\in\Zdpower{\dmn}{2n}}
\omega^{\syp{x}{y-z}}=0 \quad \mbox{whenever} \quad y\ne z,
\] 
which holds because $f_{y-z}: x\mapsto\omega^{\syp{x}{y-z}}$,
where $y\ne z$,
is a {\em character}\/ of $\Zdpower{\dmn}{2n}$ such that $f_{y-z}(x) \ne 0$
for some $x \in \Zdpower{\dmn}{2n}$ (e.g., \mycite{vanLint3rd}
or Section~III of \mycite{hamada03t}),
we obtain 
the formula (\ref{eq:twirling}), as desired.

\subsection{Twirled Channel \label{ss:tc}}

Suppose a TPCP map $\Bchgen$ on $\Bop(\Hchann)$ is given,
and the twirling is applied to the corresponding state
$\shs_{\varnn}=\hMsub{\varnn}{\Bchgen}$.
Then, the resulting state is given by (\ref{eq:twirled}), and this can be
regarded as the mixture
\[
\shs_{\varnn}'=
\frac{1}{\dmn^{2\varnn}} \sum_{x\in\myFpower{2\varnn}} \hMsub{\varnn}{\cN_x\Bchgen\cN_x^{-1}}
\]
where $\cN_{x}: \sigma \mapsto \Ebe_x \sigma \Ebe_x^{\dagger}$
and $\cM\cL$ denotes the composition that maps $\sigma$ 
to $\cM\mbm{(}\cL(\sigma)\mbm{)}$, etc.,
on account of the representation of CP maps in (\ref{eq:choiN})
[and the block structure of Choi's matrix mentioned below (\ref{eq:choiN})].
In other words, the channel $\twA$
that corresponds to the twirled state $\shs_n'$ via $\twA=\hMsubinv{\varnn}{\shs_{\varnn}'}$
is given by 
\begin{equation}\label{eq:twlchan}
\twA=\frac{1}{\dmn^{2\varnn}}\sum_{x\in\myFpower{2\varnn}}\cN_x\Bchgen\cN_x^{-1}.
\end{equation}
Since the matrix of $\hMsub{\varnn}{\twA}$ is diagonal with respect to the basis $\{ \ket{\Psi_x} \}_{x\in\myFpower{2\varnn}}$, the 
channel $\twA$ can be expressed as
\[
\twA: \sigma \mapsto \sum_{x} P_{\Bchgen}(x) \Ebe_x \sigma \Ebe_x^{\dagger},
\]
where $P_{\Bchgen}$ is 
the probability distribution on $\myFpower{2\varnn}$ defined by
\begin{equation}\label{eq:PEve}
P_{\Bchgen}(x) = \bra{\Psi_{x}} \hMsub{n}{\Bchgen} \ket{\Psi_{x}}, \quad x\in\cX^{n}
\end{equation}
with the basis $\{\ket{\Psi_{x}} \}$ in Lemma~\ref{lem:Phix}.

\section{Fidelity of Symplectic Codes \label{ss:sc}}

In this section, we present the formula for the fidelity of symplectic codes.
A self-contained exposition of symplectic codes, 
as well as proofs 
of the lemmas in this section,
can be found in \ref{app:sc},
which is a recast of Section~III of \mycite{hamada02c}
except the proof of Theorem~\ref{lem:Fengen}.

Recall that a symplectic code is obtained from
a subspace $\Cso \subset \Zdpower{\dmn}{2\genn}$ that is contained in
the symplectic dual $\Cso^{\perpsyp}$ of $\Cso$.
Specifically, (a code subspace of) a symplectic code associated with $\Cso$
is a subspace of the form
\[
\{ \psi \in \Hgem \mid \Ebe_x \psi = \tau(x) \psi, \ x \in \Cso \}
\]
where $\tau(x)$, $x \in\Cso$, are some complex numbers. 
When $\dim_{\myF}\Cso=\genn-\genk$, we have $\dmn^{\genn-\genk}$ such
subspaces, and the collection of these subspaces is also referred to
as the symplectic code associated with $\Cso$.
With a basis $(\gfL_1,\dots,\gfL_{\genn-\genk})$ of $\Cso$ fixed,
we have $\dmn^{\genn-\genk}$ cosets of $\Cso^{\perpsyp}$ in $\myFpower{2\genn}$
of the form $\{ x \in\myFpower{2\genn} \mid \syp{\gfL_i}{x} = s_i, \
i=1,\dots,\genn-\genk \}$, where $s=(s_1,\dots,s_{\genn-\genk})\in\myFpower{\genn-\genk}$.
Thus, we can label the cosets of $\Cso^{\perpsyp}$ by
$s\in\myFpower{\genn-\genk}$. It is known that there is a one-to-one correspondence between the set of
these cosets and that of the code subspaces, $\Cqarg{(s)}$, $s\in\myFpower{n-k}$.
For the specification of $\Cqarg{(s)}$, see \ref{app:sc}.
If we choose a vector $\hat{x}(s)$ from each coset $s$ of $\Cso^{\perpsyp}$
in $\Zdpower{\dmn}{2\genn}$,
and denote the set of coset representatives $\hat{x}(s)$ by $\Jsp_0$,
we have quantum codes $(\Cqarg{(s)},\Dec^{(s)})$, $s\in\myFpower{n-k}$,
where $\Dec^{(s)}:
\Bop(\Hchann)\to\Bop(\Hchann)$ is a recovery operator
designed so that the code is $\Ebe_{\Jsp}$-correcting,
$\Jsp=\Jsp_0+\Cso$.

The recovery operator can be specified by Kraus operators, 
\begin{equation}\label{eq:ro1}
\DKO^{(s)}_t=\Ebe_{\hat{x}(t)}^\dagger\Proj_{t+s}, \quad t\in\myFpower{n-k}, 
\end{equation}
where $\Proj_{t'}$ is the projection onto
the code subspace $\Cqarg{(t')}$, 
viz.,  
\begin{equation}\label{eq:ro2}
\Dec^{(s)}(\sigma)= \sum_{t\in\myFpower{n-k}} \DKO^{(s)}_t \sigma \DKO_t^{(s)\dagger}.
\end{equation}
This operation is expressed as
the measurement $\{ \Proj_{t+s} \}_t$ followed by the unitary $\Ebe_{\hat{x}(t)}^{\dagger}$.
The measurement result $t$ represents the `relative syndrome', so to speak, 
for the code $\Cqarg{(s)}$.
We denote the trace-decreasing CP map
$\sigma \mapsto \DKO^{(s)}_t \sigma \DKO_t^{(s)\dagger}$ by $\Dec^{(s,t)}$,
so that $\Dec^{(s)}=\sum_{t\in\myFpower{n-k}}\Dec^{(s,t)}$.

Let $\mms{\Cq}$ denote the projection
operator onto $\Cq$ divided by $\dim\Cq$.
The entanglement fidelity\myspcite{schumacher96} of the $\Ebe_{\Jsp}$-correcting code $\Cq$ 
used on a channel $\Bch: \Bop(\Hgem)\to \Bop(\Hgem)$, $\sigma \mapsto 
\sum_{x\in\myFpower{2\genn}} P_{\genn}(x) \Ebe_x \sigma \Ebe_x^{\dagger}$,
where $P_{\genn}$ is a probability distribution on $\cX^{\genn}$,
is given by
\begin{equation}\label{eq:FenNchan}
\Fen(\mms{\Cqarg{(s)}}, \Dec^{(s)}\Bch) =
P_{\genn}(\Jsp)=\sum_{x\in \Jsp} P_{\genn}(x)
\end{equation}
for any $s\in\myFpower{\genn-\genk}$.
This follows from a finer analysis on the entanglement fidelity
for $\Dec^{(s,t)}$, namely, from the next lemma, which is proved in \ref{subsec:tech4sc}.
\begin{lemma}\label{lem:FenNchan}
Let a subspace $\Cso\subset\myFpower{2\genn}$ which is self-orthogonal 
with respect to the symplectic form $\syp{\cdot}{\cdot}$ 
and $\hat{x}(t)$, $t\in\myFpower{n-k}$, be given as above.
Then,
\begin{equation*} 
\Fen(\mms{\Cqarg{(s)}}, \Dec^{(s,t)}\Bch) =
P_{\genn}(\hat{x}(t)+\Cso)=\sum_{x\in \hat{x}(t)+\Cso} P_{\genn}(x)
\end{equation*}
for any $s,t\in\myFpower{n-k}$ and channel $\Bch: \Bop(\Hgem)\to \Bop(\Hgem)$, $\sigma \mapsto 
\sum_{x\in\myFpower{2\genn}} P_{\genn}(x) \Ebe_x \sigma \Ebe_x^{\dagger}$.
\end{lemma}
{\em Remark.}\/
Throughout, $\Fen$ is to be understood as the unnormalized entanglement fidelity\myspcite{barnum00}.

The corresponding statement for general channels is given in the next theorem,
which will be proved in \ref{app:sc}.
\begin{theorem}\label{lem:Fengen}
Let a subspace $\Cso\subset\myFpower{2\genn}$
and $\hat{x}(t)$, $t\in\myFpower{n-k}$, be given as above.
Then, the 
symplectic codes $(\Cqarg{(s)},\Dec^{(s)}=\sum_t \Dec^{(s,t)})$
associated with $\Cso$ satisfy
\[
\frac{1}{\dmn^{\genn-\genk}} \sum_{s\in\myFpower{\genn-\genk}}
\Fen(\mms{\Cqarg{(s)}},\Dec^{(s,t)}\Bch) = \sum_{x \in \hat{x}(t)+\Cso} P_{\Bch}(x),
\]
for any $t\in\myFpower{n-k}$ and TPCP map $\Bch: \Bop(\Hgem)\to \Bop(\Hgem)$, where
$P_{\Bch}$ is associated with $\Bch$ by (\ref{eq:PEve}).
\end{theorem}
\begin{corollary} \label{coro:Fengen}
For $\Jsp=\bigcup_{t}[\hat{x}(t)+\Cso]$, 
\[
\frac{1}{\dmn^{\genn-\genk}} \sum_{s\in\myFpower{\genn-\genk}}
\Fen(\mms{\Cqarg{(s)}},\Dec^{(s)}\Bch) = \sum_{x\in\Jsp} P_{\Bch}(x).
\]
\end{corollary}

{\em Remark.}\/
That $\sum_{x\in\Jsp} P_{\Bch}(x)$ is a lower bound
to the average fidelity in Corollary~\ref{coro:Fengen}
easily follows 
from the observation of Gottesman and
Preskill\myspcite{GottesmanPreskill01}
as remarked in \mycite{hamada03s}.

\section{Fidelity of Entanglement Distillation \label{ss:distill}}

\subsection{One-Way Protocols \label{ss:oneway}}

In this section, we will consider the problem of evaluating the fidelity 
of entanglement distillation schemes
and see its close relation to quantum error-correcting codes.
Shor and Preskill described their 
famous proof of the security of the BB84 protocol
in terms of entanglement distillation.
The entanglement distillation protocol they used 
is as follows, where
as usual, the protocol is performed by Alice and Bob.
First, imagine they are given a bipartite state $\hMsubbar{n}{\Eve_n}= [\Id \tnsr \Eve_n](\ket{\bar{\Psi}}\bra{\bar{\Psi}})$,
where $\ket{\bar{\Psi}} = \dmn^{-n/2} \sum_{\vars,\varu}\ket{\dbar{\vars,\varu}}\tnsr
\ket{\bar{\vars,\varu}}$, 
$\{ \ket{\bar{\vars,\varu}} \}_{\varu}$ is
an orthonormal basis of $\codesubs{\vars}$ for each $s\in\myFpower{n-k}$,
$\{ \ket{\dbar{\vars,\varu}} \}_{s,\varu}$ is an orthonormal
basis of $\Hgen$, and
$\codesubsbar{\vars}$ is spanned by $\ket{\dbar{\vars,\varu}}$,
$u\in\myFpower{k}$, for each $s\in\myFpower{n-k}$.
Alice performs the local measurement $\{ \Pi'_{\vars} \}$
on the first half of the system, where
$\Pi'_{\vars}$ denotes the projection onto the subspace 
$\codesubsbar{\vars}$, 
and Bob performs the recovery operation for the $\Ebe_{\Jcr}$-correcting code
$\codesubs{\vars}$ knowing that Alice's measurement result is $\vars$.
Now recall the physical meaning of entanglement fidelity\myspcite{schumacher96}: 
Suppose an ideal bipartite state  
$\ket{\Phi}=\ket{\Phs}=\dmn^{-k}\sum_u\ket{\dbar{\vars,\varu}} 
\tnsr \ket{\bar{\vars,\varu}}$
is given, where
$\{ \ket{\dbar{\vars,\varu}} \}_u$ plays the role of an orthonormal basis of
the `reference' system\myspcite{schumacher96}; then,
$\Fen(\mms{\Cq}, \cB)=
\bra{\Phi} [\Id\tnsr \cB](\ket{\Phi}\bra{\Phi}) \ket{\Phi}$.
Since Alice obtains each measurement result $\vars$
with the equal probabilities and 
the resulting state is
$[\Id\tnsr \Eve_n](\ket{\Phs}\bra{\Phs})$
conditioned on this event,
the fidelity of this distillation protocol for $\hMsubbar{n}{\Eve_n}$
is exactly the same as
the average entanglement fidelity of the code $(\codesubs{\vars},\Dec^{(\vars)})$
in Corollary~\ref{coro:Fengen}.

For the security proof, the above argument is enough\myspcite{hamada03s}.
For the purposes of entanglement distillation, however, we should start with 
$\hMsub{n}{\Eve_n}= [\Id \tnsr
\Eve_n](\ket{\Psi}\bra{\Psi})$, rather than $\hMsubbar{n}{\Eve_n}$,
since in the standard setting the given bipartite states are of the form
$\rho^{\tnsr n}$, which is written in (or
reduced by twirling to) the form $\hMsub{n}{\cA^{\tnsr n}}$.
This problem is resolved upon noticing 
the relation $(\bar{U}\tnsr U)\ket{\Psi}=\ket{\Psi}$,
which holds for any unitary $U$ by (\ref{eq:me_t}), 
and the existence of the unitary $U$ that maps
$\ket{(s,u)}$, to $\ket{\bar{\vars,\varu}}$,
$(s,u)\in\myFpower{n-k}\times \myFpower{k}\simeq\myFpower{n}$.
In fact, we can choose $\bar{U}\ket{(s,u)}$
as $\ket{\dbar{\vars,\varu}}$, $(s,u)\in\myFpower{n-k}\times \myFpower{k}$
so that $\hMsubbar{n}{\Eve_n}=\hMsub{n}{\Eve_n}$.

Thus, we see the average entanglement fidelity given in Theorem~\ref{lem:Fengen}
is the fidelity of the following one-way 
entanglement distillation protocol 
for the state $\hMsub{n}{\Eve_n}$
(or for any bipartite state $\rho_n\in\Bop(\Hchann\tnsr\Hchann)$ if
the participants of the distillation protocol perform
the discrete twirling as a preprocessing). 

{\em Protocol.}\/
First, Alice performs the orthogonal measurement consisting of the projections
onto $\codesubsbar{\vars}$,
where provided Alice's measurement result is $s$,
the resulting state is
$\rho^{(s)}=[\Id \tnsr
\Eve_n](\ket{\Phs}\bra{\Phs})$.
Bob applies the recovery operator $\Dec^{(s)}$ to his system.
Alice and Bob, respectively, apply some unitaries 
$U_{\rm A}$ and $U_{\rm B}$ such that
$U_{\rm A}\ket{\dbar{\vars,\varu}}=\ket{\dbar{0^{n-k},\varu}}$ and 
$U_{\rm B}\ket{\bar{\vars,\varu}}=\ket{\bar{0^{n-k},\varu}}$.

Protocols thus obtained will be sometimes called symplectic (entanglement) distillation protocols. This class of one-way protocols are also applicable to correlated states\myspcite{hamada02m,hamada03t}.

\subsection{Two-Way Protocols \label{ss:twoway}}

Theorem~\ref{lem:Fengen} is also useful for analyses of
two-way entanglement distillation from multiple copies of 
a state $\rho$.
In this case, the corresponding channel $\cA_n$ can be written as 
$\cA^{\tnsr n}$ for some channel
$\cA: \Bop(\Hgn)\to\Bop(\Hgn)$.
For example, consider Bennett et al.'s protocol\myspcite{bennett96p}, where Alice and Bob use the symplectic code
associated with $\spn (0,1,0,1) =\{(0,0,0,0), (0,1,0,1)\}$,
where $n=2$ and $k=1$.
[This code is sometimes called
$[[2,1]]$ cat code and the core of this distillation protocol was
originally described\myspcite{bennett96p}
in terms of quantum gates as
a decoding network of the cat code was\myspcite{ss96,dss98}.] 
The protocol consists of several iterations of the two-way procedure
using $\spn (0,1,0,1)$ and a one-way entanglement distillation protocol.
The two-way procedure using $\spn (0,1,0,1)$ is not much different from
the one-way symplectic distillation protocol using it:
In each step, Alice and Bob pair up surviving states,
and for each pair they do the same measurement and unitaries
as described in Section~\ref{ss:oneway},
where in the second or further step,
the basis $\{ \ket{l}\tnsr \ket{m} \}_{l,m\in\myFpower{n}}$
is to be understood as the 
basis $\{ \ket{\dbar{0^{n-k},\varu}}\tnsr \ket{\bar{0^{n-k},\varu'}} \}_{\varu,\varu'}$
obtained newly in the previous step.
In the present case of two-way distillation, however, they retain only states 
with result $t=0$, or $t\in T$ for some fixed proper subset of $\myFpower{n-k}$
[recall $\Dec^{(s)}=\sum_t \Dec^{(s,t)}$],
and discard the rest.
Clearly, both the two-way subroutine and the final 
one-way procedure can be replaced by arbitrary ones
based on symplectic codes that are described or exemplified above\myspcite{GottesmanLo01}, though the problem of estimating the fidelity for such
schemes is non-trivial for general states, which is solved 
by Theorem~\ref{lem:Fengen}. 

\section{Exponential Convergence of Fidelity \label{ss:exp}}

Recently, a formula for the quantum capacity written with coherent information, 
which had been conjectured by several authors, 
was confirmed\myspcite{Shor02msri,devetak03}.
Regarding this topic,
from a view point of information theory or large-deviation theory,
we will consider the problem of finding
attainable speeds of convergence (exponents) of the fidelity
of quantum codes, or other similar schemes, to unity.

A memoryless quantum channel is a TPCP map
\[
\cA: \Bop(\Hchan) \to \Bop(\Hout).
\]
The term `memoryless' refers to the property that
$\cA$ acts on a density operator $\rho$ in $\Bop(\Hchan^{\tnsr \exgnn})$
as $\cA^{\tnsr \exgnn}(\rho)$.
A coding scheme or {\em code}\/ for $\cA^{\tnsr \exgnn}$ 
is a triple $(\Hbg_{\exgnn},\Ebg_{\exgnn},\Dbg_{\exgnn})$
that consists of a Hilbert space $\Hbg_{\exgnn}$, and TPCP maps
\begin{eqnarray}
\Ebg_{\exgnn}: \Bop(\Hbg_{\exgnn}) \to \Bop(\Hchan^{\tnsr \exgnn}),\\
\Dbg_{\exgnn}: \Bop(\Hout^{\tnsr \exgnn}) \to \Bop(\Hbg_{\exgnn}).
\end{eqnarray}

\begin{definition} \label{def:ach}
A number $R$ is said to be an achievable rate for $\cA$ if there exists
a sequence of codes $(\Hbg_{\exgnn},\Ebg_{\exgnn},\Dbg_{\exgnn})$ 
for $\cA^{\tnsr \exgnn}$ 
such that 
\[
\limsup_{\exgnn\to\infty} \frac{\log_{\dmn} \dim \Hbg_{\exgnn}}{\exgnn} \ge R
\]
and
\[
\lim_{\exgnn \to \infty} 
\Fen(\mms{\Hbg_{\exgnn}},\Dbg_{\exgnn} \cA^{\tnsr \exgnn} \Ebg_{\exgnn}) = 1.
\]
\closedef

\begin{definition} \label{def:QC}
The supremum of achievable rates for a memoryless channel $\cA$
is called the quantum capacity and denoted by $\Capa(\cA)$.
\closedef

{\em Remark.}\/
This definition is essentially the same as the one using the subspace fidelity
in \mycite{barnum00}, but we employ $\Fen(\mms{\Cq},\cB)$ rather than
the minimum pure-state fidelity. For the equivalence, 
see \ref{app:proofs_aux} or examine the arguments in \mycite{barnum00}.

\begin{definition}
A number $E$ 
is said to be an attainable exponent for a channel $\cA$ and a rate $R$ if there exists
a sequence of codes $(\Hbg_{\exgnn},\Ebg_{\exgnn},\Dbg_{\exgnn})$
for $\cA^{\tnsr \exgnn}$ 
such that
\[
\liminf_{\exgnn\to\infty} \frac{\log_{\dmn} \dim \Hbg_{\exgnn}}{\exgnn} \ge R
\]
and
\[
\liminf_{\exgnn \to \infty} 
- \frac{\log_{\dmn} [1-\Fen(\mms{\Hbg_{\exgnn}},\Dbg_{\exgnn} \cA^{\tnsr \exgnn}
\Ebg_{\exgnn})]}{\exgnn} \ge E.
\]
\closedef

We will prove the next theorem in what follows.
\begin{theorem}\label{th:main}
For any memoryless channel $\cA$,
and any rate $R$ smaller than $\Capa(\cA)$,
we have a positive attainable exponent.
\end{theorem}

\section{Random Coding Bound for Symplectic Codes \label{ss:rc}}

A random coding argument shows the next lemma.
In fact, the proof of the main result of \mycite{hamada01g} 
or \mycite{hamada01e} applies to this lemma
if we replace $\cX$ thereof by $\cX^{\exgenm}$. Alternatively,
the proof in \mycite{hamada02c} works if we assume the inner code
of the concatenated code thereof to be the identity map.

\begin{lemma}\label{lem:extended_alphabet}
For any positive integer $\exgenm$, number $R$, $0 \le R <1$, and memoryless channel 
$\Enr : \Bop(\Hgn^{\tnsr \exgenm}) \to \Bop(\Hgn^{\tnsr \exgenm})$,
there exists
a sequence of symplectic codes $\{ (\Cq_{\varN}\subset\Hgn^{\tnsr \exgenm\varN},\Dec_{\varN}) \}_{\varN}$
such that $\log_{\dmn}\dim \Cq_{\varN}  \ge \exgenm\varN R$, and
\begin{equation} \label{eq:exp_1st}
1-\Expe\Fen(\mms{\Cq_{\varN}}, \Dec_{\varN} \Enr^{\tnsr\varN})
\le f(\varN) \exp_{\dmn^{\exgenm}}[-\varN E_{\exgenm}(R,\Pbg)],
\end{equation}
where
\begin{equation}\label{eq:En}
E_{\exgenm}(R,\Pbg)=\min_{Q} [ D(Q||\Pbg)/\exgenm +|1-R-H(Q)/\exgenm|^+],
\end{equation}
$\exp_{b}[y]=b^y$, $f(\varN)$ is a polynomial in $\varN$,
$|y|^+=\max\{y,0 \}$,
$H(Q)=-\sum_{x}Q(x)\log_{\dmn}Q(x)$,
$D(Q||P)=\sum_{x}Q(x)\log_{\dmn}[Q(x)/P(x)]$,
and the minimum with respect to $Q$ is taken over all
probability distributions on $\myFpower{\exgenm}$.
\enlem

{\em Remarks.}\/
The symplectic code $(\Cq_{\varN}\subset\Hgn^{\tnsr
\exgenm\varN},\Dec_{\varN})$ is to be understood as 
the ensemble
$\{ (\Cq_{\varN}^{(s)},\Dec_{\varN}^{(s)}) \}_s$,
where $s$ runs through all syndromes,
and $\Expe$ denotes the expectation operation
to produce the ensemble average with respect to the uniform distribution
over all syndromes, by which Corollary~\ref{coro:Fengen} is applicable.
The statement can be strengthen to
`For any positive integer $\exgenm$, number $R$, $0 \le R <1$, 
there exists
a sequence of symplectic codes $\{ (\Cq_{\varN},\Dec_{\varN}) \}_{\varN}$
such that $\log_{\dmn}\dim \Cq_{\varN}  \ge \exgenm\varN R$ and
for any memoryless channel 
$\Enr : \Bop(\Hgn^{\tnsr \exgenm}) \to \Bop(\Hgn^{\tnsr \exgenm})$,
(\ref{eq:exp_1st}) is satisfied'. 
This means that we can find symplectic codes whose structures do not depend 
on the channel characteristics, especially on $\Pbg$.
The proof of this refinement 
is essentially the same as that in \mycite{hamada03s}.
The proof uses the existence of a symplectic code
whose `type spectrum', which is 
a natural generalization of the weight spectrum (distribution)
in coding theory,
is `well balanced', 
and the fact that the fidelity of any symplectic code
on a memoryless channel is invariant under permutations of the coordinates (digits).

\section{Proof of Theorem~\ref{th:main}\label{ss:proofexp}}

Suppose a rate $\Rin$ is achievable for $\cA$.
Then, there exists
a sequence of codes $\{ (\Hbg_{\exgnn},\Ebg_{\exgnn}=\Ebg,\Dbg_{\exgnn}=\Dbg)
\}$ whose rate, as $\exgnn$ becomes large, approaches $\Rin$, 
which may be arbitrarily close to $\Capa(\cA)$.
We may assume
$\dim \Hbg_{\exgnn} = \dmn^{\exgenm}$ for some integer $\exgenm$ for every $\exgnn$
as argued in \ref{app:proofs_aux} (since $\Capa=\QCed$).
We apply Lemma~\ref{lem:extended_alphabet}
setting $\Enr=\Dbg\cA^{\tnsr \varm}\Ebg$ and
identifying $\Hch^{\tnsr \exgenm}$ with $\Hbg_{\varm}$.
Namely, we use two-stage coding in which 
the $n\nu$-quantum-digits system is divided into $\nu$ blocks of length $n$,
each block is coded with 
$(\Hbg_{\exgnn},\Ebg_{\exgnn}=\Ebg,\Dbg_{\exgnn}=\Dbg)$,
and $\nu$ blocks are coded with 
the codes for $\cB^{\nu}$ the existence of which is ensured
in Lemma~\ref{lem:extended_alphabet} (Fig.~\ref{fig:1}). 
The two-stage codes have overall rates not smaller than
\[
    \frac{\exgenm}{\exgnn} R.
\]

\begin{figure}[t]
\begin{center}
\begin{picture}(200,110)(0,-72)

\put(10,-74){\framebox(180,20){}}
\put(11,-73){\framebox(178,18){$\cC_{\nu}$}}

\put(25,-54){\vector(0,1){12}}
\put(75,-54){\vector(0,1){12}}
\put(175,-54){\vector(0,1){12}}

\put(2,-42){\framebox(44,14){$C_n$}}
\put(52,-42){\framebox(44,14){$C_n$}}
\put(120,-36){\makebox(0,0){$\cdots$}}
\put(152,-42){\framebox(44,14){$C_n$}}

\put(25,-28){\vector(0,1){12}}
\put(75,-28){\vector(0,1){12}}
\put(175,-28){\vector(0,1){12}}

\put(25,-9){\oval(48,14)\makebox(0,0){$\cD\cA^{\tnsr n}\cE$}}
\put(75,-9){\oval(48,14)\makebox(0,0){$\cD\cA^{\tnsr n}\cE$}}
\put(120,-9){\makebox(0,0){$\cdots$}}
\put(175,-9){\oval(48,14)\makebox(0,0){$\cD\cA^{\tnsr n}\cE$}}

\put(25,-2){\vector(0,1){12}}
\put(75,-2){\vector(0,1){12}}
\put(175,-2){\vector(0,1){12}}

\put(100,20){\oval(198,20)}
\put(100,20){\oval(196,18)\makebox(0,0){$\cR_{\nu}$}}

\end{picture}
\end{center}
\caption{The two-stage code in the proof of Theorem~\protect\ref{th:main}
consisting of the inner code $(C_n,\cE,\cD)$ 
and the outer (symplectic) code 
$(\cC_{\nu},\cR_{\nu})$, where $C_n \simeq \Hgn^{\tnsr m}$ and $\cA$
is a memoryless channel.
\label{fig:1}}
\end{figure}

From (\ref{eq:En}),
$E_{\exgenm}(R,\Pbg)$ is positive if $R<1-H(P_{\Enr})/\exgenm$, i.e., if
\begin{equation}\label{eq:threshR}
    \frac{\exgenm}{\exgnn} R<\frac{\exgenm}{\exgnn}\bigg[1-\frac{H(P_{\Enr})}{\exgenm}\bigg].
\end{equation}
The number 
$1-H(P_{\Enr})/\exgenm$ can be bounded as
\[
1-\frac{H(P_{\Enr})}{\exgenm} \ge 1 - \frac{h(P_{\Enr}(0^{2\exgenm})) + [1-P_{\Enr}(0^{2\exgenm})] 2\exgenm }{\exgenm} ,
\]
where $h$ is the binary entropy function. 
Note also by the definition of the entanglement fidelity, we have
\[
\Fen(\mms{\Hbg_{\exgnn}},\Dbg\cA^{\tnsr \varm}\Ebg)
=\Fen(\mms{\Hbg_{\exgnn}},\Enr)
=\bra{\Psi}\hMsub{\exgenm}{\Enr}\ket{\Psi} = P_{\Enr}(0^{2\exgenm}),
\]
where $\ket{\Psi}=\ket{\Psi_{0^{2m}}}$.
Then, because $\exgenm/\exgnn$ and $\Fen(\mms{\Hbg_{\exgnn}},\Dbg\cA^{\tnsr \varm}\Ebg)=P_{\Enr}(0^{2\exgenm})$
tend to $r$ and 1, respectively, as $\exgnn$ grows large, 
the number on the right-hand side of (\ref{eq:threshR}), 
for a large enough $\exgnn$, will be
arbitrarily close to $\Rin$, which in turn can be made close to $\Capa(\cA)$.

Thus, we have a sequence of 
codes of desired performance for $\cA^{\varM}$, $\varM=\varm, 2\varm,\dots$.
To interpolate a code for $\cA^{\varM}$ with $\varM=\varm\varN+i$, $0<i<\varm$, 
into this sequence,
we just past a trivial code of dimension one for the $i$-quantum-digit 
system to the large code for $\exgnn\varN$-quantum-digit system.

{\em Remark.}\/
Instead of
assuming $\dim \Hbg_{\exgnn} = \dmn^{\exgenm}$ for some integer $\exgenm$
to use the argument in \ref{app:proofs_aux},
we can generalize Lemma~\ref{lem:extended_alphabet} so that
it applies to memoryless channels $\Enr : \Bop(\Hgn') \to \Bop(\Hgn')$ with $\dim\Hgn'$ arbitrary but finite.
To do this, write the number $\dim\Hgn'$ as the product of the prime factors 
$\dmn_1\cdots\dmn_{\exgenm}$, and use the tensor product
of code subspaces of symplectic codes for 
quantum-$\dmn_i$-ary-digit systems.

\section{Exponents for Entanglement Distillation \label{ss:exp_distill}}

The above argument can be 
accommodated to the problem of
entanglement distillation from multiple copies of
a bipartite state\myspcite{bennett96p,bennett96m,DevetakWinter03d}.
In fact, the achievability of a rate, the capacity analog $\Dopt_{C}(\rho)$
(sometimes called the distillable entanglement),
and attainable error exponents for a bipartite state $\rho$ 
can be similarly defined for a given class of distillation protocols $C$ \myspcite{hamada03t}. Assume that
the participants of a protocol 
are allowed to
apply a one-way symplectic distillation protocol 
to multiple copies of $\cD(\rho^{\tnsr\varm})$ in the class $C$, where
$\cD$ is another protocol in $C$.
Note that most of protocol classes discussed in the literature,
e.g., $C_1$ through $C_{\Gamma}$ of \mycite{rains01}, possess
this property.
Then, since
the symplectic quantum code in Lemma~\ref{lem:extended_alphabet}
can be used as a one-way symplectic distillation protocol,
we conclude that 
for any state $\rho$ of a bipartite system 
and any rate below $\Dopt_{C}(\rho)$, 
we have a positive attainable exponent.

Clearly, this conclusion as well as its reasoning
extends to the scenario of entanglement generation over memoryless quantum channels, where the sender Alice 
begins with an arbitrary initial bipartite state 
$\rho_{\rm in}$ in some prescribed class $C_{\rm in}$, 
sends the half of $\rho_{\rm in}$ to produce 
$[\Id \tnsr \cA^{\tnsr n}]( \rho_{\rm in})$,
and then Alice and Bob apply some distillation protocol
to $[\Id \tnsr \cA^{\tnsr n}]( \rho_{\rm in})$ 
allowed in a prescribed class $C$.
[The term `entanglement generation' is from \mycite{devetak03}, 
where the allowed operations are those of 
local TPCP maps at the receiver's end.]

\section{Conclusion \label{ss:conc}}

In summary, based on Weyl's ray representation of $(\SINT/\dmn\SINT)^{2n}$,
with which the standard symplectic form was associated
naturally in considering the commutation relation for the representation,
the fidelities of 
schemes for quantum information processing using the property of 
the symplectic geometry were evaluated.

\section*{Acknowledgments}

The author is grateful to H.~Imai of the QCI project for support.

\appendix

\section{Basics of Symplectic Codes \label{app:sc}}

\mysubsectionapp{Symplectic Codes \label{subsec:sc}}

In this section, the framework of symplectic codes is rebuilt
on the theory of geometric algebra\myspcite{artin,grove}.
For a subspace $\Cso\in \myFpower{2n}$, let $\Cso^{\perp}$ be defined by
\[
\Cso^{\perp} = \{ y\in\myFpower{2n} \mid \forall x\in\Cso,\ \syp{x}{y}=0 \}.
\]
By linear algebra, the matrices of commuting unitary operators
are diagonal with respect to a common basis.
A symplectic code is a collection of simultaneous eigenspaces 
of a set of commuting operators in the Weyl basis.
By (\ref{eq:wcr}),
if a set $\Cso\subset\myFpower{2n}$ has the property that the operators
$\Ebe_x$, $x\in\Cso$, commute with each other, then $\spn \Cso$
has the same property. Hence, it is enough to consider
a subspace $\Cso\subset\myFpower{2n}$ such that
\[
\forall x,y\in\Cso,\quad \syp{x}{y}=0,
\]
which is equivalent to $\Cso\subset \Cso^{\perp}$.
A subspace $\Cso \in \myFpower{2n}$ 
is said to be {\em self-orthogonal}\/ (with respect to the symplectic bilinear form)
if $\Cso\subset \Cso^{\perp}$.

The statement of
the following lemma can be found in
\mycite{gottesmanPhD}, Section~3.2, and \mycite{gottesman99}.
A proof based on
the very basics of symplectic geometry\myspcite{artin,grove}
has been given in \mycite{hamada02c}.

\begin{proposition}\label{lem:hyperbolic_plane}
Let $\Cso$ be a self-orthogonal subspace with 
$\dim \Cso = n-k$ and $\Cso= \spn \{ g_1, \dots , g_{n-k} \}$. 
Then, we can find vectors $g_{n-k+1},\dots,g_n$ and $h_1,\dots,h_n$
such that 
\begin{equation}\label{eq:hyperbolic_plane}
\begin{array}{lll}
\syp{g_i}{h_j} &=& \delta_{ij}, \\
\syp{g_i}{g_j} &=& 0,\\
\syp{h_i}{h_j} &=& 0
\end{array}
\end{equation}
for $i,j=1,\dots,n$, where $\delta_{ij}$ is the Kronecker delta.
\enprop

A pair of
linearly independent vectors $(g,h)$ with $\syp{g}{h}=1$ is called
a {\em hyperbolic pair}\/, and it is known that
a space with a nondegenerate symplectic form, such as the one
defined by (\ref{eq:syp}),
can be decomposed into an orthogonal sum of the form 
\[
\spn \{w_1,z_1\} \perp \dots \perp \spn \{w_n,z_n\}
\]
in such a way that
$(w_i,z_i)$, $i=1,\dots,n$, are hyperbolic pairs\myspcite{artin}.
Following Artin\myspcite{artin}, we have referred 
to the direct sum of $U_1,\dots,U_n$ as the orthogonal sum of spaces $U_1,\dots,U_n$
if $U_1,\dots,U_n$ are orthogonal. 
The three equations in the above lemma say
that $\myFpower{2n}$ is the orthogonal sum of $\spn \{ g_i,h_i \}$,
$i=1,\dots,n$.
In the present case with the bilinear form in (\ref{eq:syp}),
the simplest example of such a decomposition of the space $\myFpower{2n}$
is $\spn \{ e_1,e_2 \} \perp \dots \perp \spn \{ e_{2n-1},e_{2n} \}$,
where $\{ e_i \}_{1\le i \le 2n}$ is the standard basis of $\myFpower{2n}$
that consists
of $e_i=(\delta_{ij})_{ 0 \le j \le 2n}\in\myFpower{2n}$, $1\le i \le 2n$.

{\em For the remainder of this appendix, we fix an arbitrary self-orthogonal 
subspace $\Cso$ with $\dim \Cso = n-k$ 
and such hyperbolic pairs $(g_1,h_1), \dots, (g_{n},h_{n})$ 
as given in Proposition~\ref{lem:hyperbolic_plane}.}\/
Any vector $x \in \myFpower{2\varnin}$ can be expanded into
\begin{equation}\label{eq:expand_x}
 x = \sum_{i=1}^{n} (w_i g_i+ z_i h_i).
\end{equation}
Thus, the hyperbolic pairs 
$(g_1,h_1), \dots, (g_{n},h_{n})$ determines the map that sends $x$ to
 $(w_1,z_1,\dots,w_{\varnin},z_{\varnin})$, which is clearly an isometry.
For $z=(z_1,\dots,z_m)\in\myFpower{m}$, $1 \le m \le n$, we write
\begin{equation}\label{eq:only_h}
\Ebeh{z}=\prod_{i=1}^{m} (\Ebe_{h_i})^{z_i} 
\end{equation}
where the product on the right-hand side is unambiguous because
$(\Ebe_{h_i})^{z_i}$, $i=1,\dots,m$, commute with each other.
Note that by (\ref{eq:prW}), 
$\Ebeh{z}$ and $\Ebe_{x}$, where $x={\dss \sum_{i=1}^{m} z_i h_i}$,
are the same up to a phase factor.
Similarly, for $w=(w_1,\dots,w_m)\in\myFpower{m}$, $1 \le m \le n$,
we write
\begin{equation}\label{eq:only_g}
\Ebeg{w}=\prod_{i=1}^{m} (\Ebe_{g_i})^{w_i}. 
\end{equation}
We have seen that any basis $\{ g_1,\dots,g_{n-k} \}$
of a self-orthogonal space can be extended to $\{ g_1,\dots,g_n \}$
in such a way that $\spn \{ g_1,\dots,g_n \}$ is self-orthogonal.
Since $\Ebe_{g_i}$, $i=1,\dots,n$, commute with each other,
we can find a basis of $\Bop(\Hch)$ on which $\Ebe_{g_i}$
are simultaneously diagonalized in matrix forms.
Hence, we can find an $n$-tuple of scalars
$( \mu_{i} )_{1\le i\le n}$ for which the space consisting of $\psi$ with
\begin{equation}\label{eq:eigenspace}
\Ebe_{g_i} \psi = \mu_i \psi, \quad i=1,\dots,n,
\end{equation}
is not empty. We call a nonzero vector (respectively, the set of vectors) 
satisfying (\ref{eq:eigenspace})
an eigenvector (respectively, the eigenspace) 
of $\{ \Ebe_{g_i} \}_{1\le i\le n}$ 
with eigenvalue list $( \mu_{i} )_{1\le i\le n}$.
Take a normalized vector 
$\ket{\ghb{0,\dots,0}}$ from this eigenspace, 
where the label $(0,\dots,0)$ belongs to $\myFpower{n}$.
Applying an operator $\Ebe_{x}$ to both sides of (\ref{eq:eigenspace})
from left
and using (\ref{eq:wcr}) as well as the symplectic property
\[
\syp{x}{y}=-\syp{y}{x},
\]
we have
\[
\Ebe_{x} \Ebe_{g_i} \psi = \mu_i \Ebe_{x} \psi,
\]
that is,
\begin{equation}\label{eq:sc_property}
\Ebe_{g_i} \Ebe_{x} \psi = \mu_i \omega^{\syp{g_i}{x}}
\Ebe_{x} \psi.
\end{equation}
This means that $\Ebe_{x}\psi$ is an eigenvector with
eigenvalue list $( \mu_{i} \omega^{\syp{g_i}{x}} )_{1\le i\le n}$.
If we expand $x$ 
as in (\ref{eq:expand_x}),
then we have $\syp{g_i}{x} = z_i$, $i=1,\dots,n$, and hence
there are, at least, $\dmn^n$ possible eigenvalue lists
for $\{ \Ebe_{g_i} \}_{1\le i \le n}$.
However, for any pair of distinct eigenvalue lists, the corresponding
eigenspaces of $\{ \Ebe_{g_i} \}_{1 \le i \le n}$ are orthogonal,
and hence there are no more eigenvalue lists.
Thus, we have an orthonormal basis
$\{\ket{\ghb{s_1,\dots,s_n}}\}_{(s_1,\dots,s_n)\in\myFpower{n}}$ defined by
\begin{equation}\label{eq:basis4Hn}
\ket{\ghb{s_1,\dots,s_n}} = \Ebeh{s} \ket{\ghb{0,\dots,0}},\quad
\mbox{where} \quad {\dss s=(s_1,\dots,s_n)}.
\end{equation}
It is easy to check that $(\Ebe_{(\varss,\vartt)})^\dmn$ is the identity
operator,
which implies eigenvalues of $\Ebe_x$, $x\in\myFpower{2n}$,
are $\dmn$-th roots of unity. 
Hence, we can take $\mu_i$, $1\le i \le n$, to be all one,
which we will assume throughout.
Note that the basis
$\{\ket{\ghb{s_1,\dots,s_n}}\}_{(s_1,\dots,s_n)\in\myFpower{n}}$
depends on $(g_i,h_i)$, $i=1,\dots,n$.

We expand $x$ as in (\ref{eq:expand_x}) and put
\begin{equation}
\begin{array}{lll}
\newx &=& (z_1,\dots,z_n), \\ 
\newz &=& (w_1,\dots,w_n).
\end{array}\label{eq:te0}
\end{equation}
Define $[a,b]$ as $(a_1,b_1,\dots,a_n,b_n)\in\myFpower{2n}$,
$X^{a}$ as $X^{\oldx_1}\tnsr\cdots\tnsr X^{\oldx_n}$ and
$Z^{b}$ as $Z^{\oldz_1}\tnsr\cdots\tnsr Z^{\oldz_n}$
for $a=(\oldx_1,\dots,\oldx_n),b=(\oldz_1,\dots,\oldz_n)\in\myFpower{n}$.
Then, 
$\Ebe_{[a,b]}=X^aZ^b$,
\[
X^{\oldx}\ket{\varb_{1},\dots,\varb_n}
=\ket{\varb_{1}-\oldx_1,\dots,\varb_n-\oldx_n}
\]
and
\[
Z^{\oldz}\ket{\varb_{1},\dots,\varb_n}
=\prod_{i=1}^{n}\omega^{\oldz_i \varb_i}\ket{\varb_{1},\dots,\varb_n},
\]
$a,b,(l_1,\dots,l_n)\in\myFpower{n}$,
by the definitions of $\Ebe$, $X$ and $Z$.
We notice that the actions of 
$\newX{\newx}$ and $\newZ{\newz}$,
$\newx,\newz\in\myFpower{n}$, on the new basis is quite similar to 
those of $X^{-1}$ and $Z$ on $\ket{\varb_{1},\dots,\varb_n}$:
\begin{equation}\label{eq:barX}
\newX{\newx}\ket{\ghb{\varb_{1},\dots,\varb_n}}
=\ket{\ghb{\varb_{1}+\newx_1,\dots,\varb_n+\newx_n}}
\end{equation}
and
\begin{equation}\label{eq:barZ}
\newZ{\newz}\ket{\ghb{\varb_{1},\dots,\varb_n}}
=\prod_{i=1}^{n}\omega^{\newz_i \varb_i}\ket{\ghb{\varb_{1},\dots,\varb_n}},
\end{equation}
$\newx,\newz,(\varb_{1},\dots,\varb_n)\in\myFpower{n}$.
Eq.~(\ref{eq:barX}) holds by definition, and (\ref{eq:barZ}) can be checked
as follows.

{\em Proof of (\ref{eq:barZ}).}\/
Since $\Ebe$ is a ray representation,
$\newX{l}$ and $\newZ{\newz}$ can be written as
\[
\newX{l}=\lambda\Ebe_{\Sigma_{i}l_i h_i},
\newZ{\newz}=\lambda'\Ebe_{\Sigma_{i}\newz_i g_i}
\]
with some constants $\lambda$ and $\lambda'$,
where $l=(l_1,\dots,l_n)$ and $i$ runs through 1 to $n$ in the summations.
Then,
\begin{eqnarray*}
\newZ{\newz}\ket{\bar{l}} &=& \newZ{\newz}\newX{l}\ket{\bar{0^n}}\\
&=& \lambda\lambda' \Ebe_{\Sigma_{i}\newz_i g_i}\Ebe_{\Sigma_{i}l_i h_i}\ket{\bar{0^n}}\\
&\stackrel{\rm (a)}{=}& \lambda\lambda' \omega^{\syp{\Sigma_{i}\newz_i g_i}{\Sigma_{i}l_i h_i}}\Ebe_{\Sigma_{i}l_i h_i}\Ebe_{\Sigma_{i}\newz_i g_i}\ket{\bar{0^n}}\\
&=& \omega^{\Sigma_{i}\newz_i l_i} \newX{l}\newZ{\newz}\ket{\bar{0^n}}\\
&\stackrel{\rm (b)}{=}& \omega^{\Sigma_{i}\newz_i l_i} \newX{l}\ket{\bar{0^n}}
=\omega^{\Sigma_{i}\newz_i l_i} \ket{\bar{l}},
\end{eqnarray*}
where the equalities (a) and (b) follow from (\ref{eq:wcr}) and
(\ref{eq:eigenspace})
with the assumption $\mu_i=1$ for all $i$, respectively.
\enproof

Now we are ready to see
the principle of symplectic codes.

\begin{proposition}\cite{crss97,crss98,gottesman96}. \label{lem:coset_leaders}
Let a subspace $\Cso\subset\myFpower{2n}$ satisfy 
\begin{equation}\label{eq:self-orth}
\Cso\subset \Cso^{\perp} \quad \mbox{and} \quad \dim \Cso = n-k.
\end{equation}
In addition, let $\Icr_0 \subset \myFpower{2n}$ 
be a set satisfying
\begin{equation} 
\forall x,y\in \Icr_0,\ [ \, y-x \in \Cso^{\perp} \Rightarrow x=y \, ],
\end{equation}
and put
\[
 \Icr=\Icr_0+\Cso=\{ z+w \mid z\in\Icr_0,w\in\Cso \}.
\]
Then, the $\dmn^{k}$-dimensional subspaces of the form
\begin{equation}\label{eq:codespace}
\{ \psi \in \Hch^{\tnsr n} \mid \forall M\in\Ebe_{\Cso},\ M \psi =  \tau(M) \psi \},
\end{equation}
where $\tau(M)$ are eigenvalues of $M\in \Ebe_{\Cso}$,
are $\Ebe_{\Icr}$-correcting codes.
\enprop

In fact, the subspace 
\begin{equation}\label{eq:codespace2}
\Hcd^{(s)} = \spn \{ \ket{\ghb{\varsp_1,\dots,\varsp_{n-k},\varsp_{n-k+1},\dots,\varsp_n}} \mid (\varsp_{n-k+1},\dots,\varsp_n) \in \myFpower{k} \}
\end{equation}
with a fixed $(n-k)$-tuple $s=(\varsp_1,\dots,\varsp_{n-k})\in\myFpower{n-k}$ 
is such a quantum code.
The equivalence of (\ref{eq:codespace}) and (\ref{eq:codespace2})
follows from (\ref{eq:prW}).
Since there are $\dmn^{n-k}$ possible choices for $(\varsp_1,\dots,\varsp_{n-k})$,
we have $\dmn^{n-k}$ codes.
The term {\em codes}\/ is applied to both a self-orthogonal subspace
$\Cso \subset \myFpower{2n}$, and quantum codes $\Hcd^{(s)}$
associated with $\Cso$. The collection of quantum codes $\Hcd^{(s)}$
or one from the collection, each possibly accompanied by a recovery operator,
is called a {\em symplectic code associated with $\Cso$ or symplectic
(stabilizer) code with stabilizer $\Ebe_{\Cso}$}.

Since $\Cso^{\perp}$ is spanned by $g_1,\dots,g_n$ and $h_{n-k+1},\dots,h_n$,
any coset of $\Cso^{\perp}$ in $\myFpower{2n}$ is of the form
\begin{eqnarray}
\lefteqn{\Big\{   \sum_{i=1}^{n} (w_i g_i+ z_i h_i) 
\mid z_i = s_i,\, i=1,\dots,n-k \Big\}}\nonumber\\
&=&\{ x \mid \syp{g_i}{x} = s_i,\, i=1,\dots,n-k  \}\label{eq:coset}
\end{eqnarray}
with some $(n-k)$-tuple $s=(s_1,\dots,s_{n-k})$.
The set of cosets of $\Cso^{\perp}$ and 
$\{ \Ebe_{\varrc} \Hcd^{(0)} \mid \varrc \in \Icr_0 \}$,
where $\Ebe_{\varrc} \Hcd^{(0)}$ denotes $\{ \Ebe_{\varrc}\psi \mid
\psi\in\Hcd^{(0)} \}$ with $0$ being the abbreviation of $(0,\dots,0)\in\myFpower{n-k}$,
are in a one-to-one correspondence
when $\Icr_0$ is a transversal 
(a set of coset representatives
such that each coset has exactly one representative in it), 
i.e., when $\crd{\Icr_0}=\dmn^{n-k}$.
In fact, for any vector $x$ in the coset in (\ref{eq:coset}),
we have, 
by (\ref{eq:barX}) and (\ref{eq:barZ}) or Section~A.3 below, 
\begin{equation}\label{eq:synd2}
\Hcd^{(s)}= \Ebe_{\varrc} \Hcd^{(0)}.
\end{equation}
The $(n-k)$-tuple $(s_i)_{1 \le i \le n-k}$ is called a syndrome on the analogy
with classical linear codes. 

To show that the subspace, say $\Hcd$, 
in (\ref{eq:codespace}) or (\ref{eq:codespace2})
is really $\Ebe_{\Icr}$-correcting, we may use 
Theorem III.2 of Knill and Laflamme\myspcite{KnillLaflamme97}.
Alternatively, we can directly check the error-correcting capability
using the recovery operator specified by (\ref{eq:ro1}) and (\ref{eq:ro2})
as will be done in Section~A.3. 

\mysubsectionapp{Coset Arrays \label{ss:ca}}

In discussing symplectic codes,
it is often useful to conceive 
a {\em coset array}\/ of $\Cso$ which has the form
\begin{equation}\label{eq:carray1}
\begin{array}{rrrr}
y_0+x_0+\Cso & y_0+x_1 +\Cso  & \cdots & y_0+x_{\varDK-1} +\Cso \\
y_1+x_0+\Cso & y_1 + x_1 +\Cso  & \cdots & y_1+x_{\varDK-1} +\Cso \\
\vdots\ & \vdots\ & & \vdots\ \\
y_{\varNK-1}+x_0+\Cso & y_{\varNK-1}+ x_1 +\Cso  & \cdots & y_{\varNK-1}+x_{\varDK-1} +\Cso
\end{array}
\end{equation}
where $\varDK=\dmn^{2k}$, $\varNK=\dmn^{n-k}$,
$\{ x_i \}$ is a transversal 
of the cosets of $\Cso$ in $\Cso^{\perp}$,
and $\{ y_i \} $ is that of the cosets of $\Cso^{\perp}$ in $\myFpower{2n}$.
Here, the integer index $i$ of $x_i$
is identified with $s\in\myFpower{n-k}$,
which can be viewed as a $\dmn$-ary number, 
and that of $y_i$ is to be similarly understood.
In the array, each entry is a coset of $\Cso$ in $\myFpower{2n}$,
and each row form a coset of $\Cso^{\perp}$ in $\myFpower{2n}$.
This array resembles standard arrays
often used in classical coding theory\myspcite{slepian56,ptrsn}, 
and there is an analogy between them.
For example, if we choose one coset $y_{s}+x_{u}+\Cso$ from each row,
and denote the union of these cosets by $\Jcr$, then
there are recovery operators such that
the resulting symplectic codes are $\Ebe_{\Jcr}$-correcting, 
which was already mentioned in the previous section and will be proved
in the next section.
[Entries of a standard array of a classical linear code
are not cosets but vectors, and if we choose a vector from each row,
and denote the set of these vectors by $\Jcr$, then
we can decode it in such a way that
the resulting code is $\Jcr$-correcting.]

\mysubsectionapp{Proof of Lemma~\ref{lem:FenNchan}: Fidelity of Codes on 
Channels Subject to Probabilistic Weyl Unitaries 
 \label{subsec:tech4sc}}

To calculate the fidelity, we trace the action of $I\tnsr\Ebe_{x}$
on the state $\ket{\Phi_s}\bra{\Phi_s}$, where
\[
\ket{\Phi_s} = \frac{1}{\dmn^{k/2}} \sum_{(\varb_1,\dots,\varb_k)\in\myFpower{k} }
\ket{\ghb{\varb_{1},\dots,\varb_k}} 
\tnsr \ket{\ghb{s_1,\dots,s_{n-k},\varb_{1},\dots,\varb_k}}
\]
is a purification of $\mms{\Hcd^{(s)}}$, $s=(s_1,\dots,s_{n-k})$.

Suppose an error $\Ebe_x$, $x\in\myFpower{2n}$,
has occurred on a state $\mms{\Hcd^{(s)}}$.
We decompose $x$ into
\[
x=\sum_{i=1}^{n-k} w_i g_i +\sum_{i=1}^{n-k} z_i h_i + \sum_{i=1}^{k} z_{i+n-k} h_{i+n-k}+\sum_{i=1}^{k} w_{i+n-k} g_{i+n-k}.
\]
Then, $\Ebe_x$ is the same as $U_3U_2U_1$ up to an irrelevant phase factor,
where $U_1=\newZ{v}$, $v=(w_1,\dots,w_{n-k})$,
$U_2=\newX{t}$, $t=(z_1,\dots,z_{n-k})$, 
and $U_3=\newX{u}\newZ{u'}$, $u=(0,\dots,0,z_{n-k+1},\dots,z_{n})$,
$u'=(0,\dots,0,w_{n-k+1},\dots,w_{n})$.
By (\ref{eq:barX}) and (\ref{eq:barZ}), 
$I\tnsr(U_2U_1) \ket{\Phi_s}\bra{\Phi_s}  I\tnsr(U_2U_1)^{\dagger} =  I\tnsr U_2 \ket{\Phi_s}\bra{\Phi_s}  I\tnsr U_2^{\dagger} = 
\ket{\Phi_{s+t}}\bra{\Phi_{s+t}}$.
The final part $I\tnsr U_3$ acts
on the state $\ket{\Phi_{s+t}}\bra{\Phi_{s+t}}$ as a Weyl unitary.
[These actions may be visualized in terms of a coset array as follows.
Assume for simplicity $s=0^{n-k}$,
recall $\Ebe_{x}\Cqarg{(0)}=\Cqarg{(t)}$, and write $\Cqarg{(i)}$ 
beside 
the $i$-th row of the array;
$U_1$ does nothing, $U_2$ translates the half of the state 
$\ket{\Phi_0}$ along the vertical lines to $\Cqarg{(t)}$ 
and $U_3$ acts as the Weyl unitary 
specified by $(u,u')$ that corresponds to a horizontal index of the array
in a one-to-one fashion.]

Now suppose $\hat{x}(t)$ is expanded as $x$ was to yield $\hat{v}$,
$\hat{u}$ and $\hat{u}'$ in place of $v$, $u$ and $u'$.
Then, only the effect of errors $\Ebe_x$ such that $u=\hat{u}$ and
$u'=\hat{u}'$ is properly canceled out
by applying $\Ebe_{\hat{x}(t)}^{\dagger}$.
In fact, by Lemma~\ref{lem:Phix}, 
the entanglement fidelity equals one if 
$x\in \hat{x}(t)+\Cso$
and zero otherwise
since the final states
is $\bar{X}^{u-\hat{u}}\bar{Z}^{u'-\hat{u}'}\ket{\Phi_s}$.
Hence, we obtain the desired formula.

\mysubsectionapp{Proof of Theorem~\ref{lem:Fengen}}

Suppose the twirling is applied to $\shs_{\genn}=\hMsub{\genn}{\Bch}$
to yield $\hMsub{\genn}{\twA}$.
Since the matrix of $\hMsub{\genn}{\twA}$ is diagonal with respect to the basis $\{ \ket{\Psi_x} \}_{x\in\myFpower{2\genn}}$, the mixed channel $\twA$ has the form 
$\twA: \sigma \mapsto \sum_{x} P_{\genn}(x) \Ebe_x \sigma \Ebe_x^{\dagger}$ with the
probability distribution $P_{\genn}=P_{\Bch}$ on $\myFpower{2\genn}$
by Theorem 1 of Choi\myspcite{choi75} as argued in Section~\ref{ss:tc}.

Now assume $\Cq\subset\Hgem$ is a code subspace, say $\Cqarg{(0)}$,
of the symplectic code. 
Then, by Lemma~\ref{lem:FenNchan}
\begin{eqnarray*}
P_{\Bch}(\hat{x}(t)+\Cso)&=&\Fen(\mms{\Cq},\Dec^{(0,t)}\twA)\\
&=&\Fen(\mms{\Cq},\dmn^{-2\genn}\Dec^{(0,t)}\sum_{x}\cN_x\Bch\cN_x^{-1})\\
&=&\frac{1}{\dmn^{2\genn}}\sum_{x}\Fen(\cN_x^{-1}(\mms{\Cq}),\cN_x^{-1}\Dec^{(0,t)}\cN_x\Bch).
\end{eqnarray*}
Since $N_x^{\dagger}\Cq = \{ N_x^{\dagger} \psi \mid \psi \in\Cq \}$
ranges uniformly over the whole set of code subspaces of the symplectic code
associated with $\Cso$ (Section~A.3 of this appendix 
or Section~III of \mycite{hamada02c})
as $x$ runs through $\myFpower{2\genn}$
[and $\cN_x^{-1}\Dec^{(0,t)}\cN_x=\Dec^{(s,t)}$ as can be checked easily],
this means that the entanglement fidelity 
$\Fen(\mms{\Cqarg{(s)}},\Dec^{(s,t)}\Bch)$ of the symplectic code
averaged over all code subspaces $\Cqarg{(s)}$,
$s\in\myFpower{\genn-\genk}$, is given by
$P_{\Bch}(\hat{x}(t)+\Cso)$, as promised.

 \mysubsectionapp{Remark on Symplectic Stabilizer Codes \label{subsec:remark4sc}}

If we define $\Ebe$ by (\ref{eq:error_basis_d3})
for $\dmn=2$, most existing arguments on symplectic codes work.
In this case, however, we cannot assume 
$\mu_i$, $1\le i \le n$, to be all one in general.
For example, recall the eigenvalues of $XZ$.

\section{Fidelities and Quantum Capacity \label{app:proofs_aux}}

In this appendix, 
only for a technical reason, we define three variants of $\Capa$,
which will appear as $\QCed$, $\QCp$ and $\QCpd$, and
show that these are all equal to each other.
This fact is used in the proof of Theorem~\ref{th:main}
in Section~\ref{ss:proofexp}.

In Definition~\ref{def:ach}, we could have used minimum pure state fidelity
\[
F_{\rm p}(\Hbg,\Enr) = \min_{\ket{\phi}\in\Hbg:\, \|\phi \|=1} \bra{\phi} \Enr(\ket{\phi}\bra{\phi})\ket{\phi}
\]
in place of entanglement fidelity. The $\QCp$ is defined 
in the same way as $\Capa$ with $\Fen$ replaced by $F_{\rm p}$.
In Definition~\ref{def:ach}, we could also have restrict ourselves to codes
$\{ (\Hbg_{\exgnn},\Ebg_{\exgnn},\Dbg_{\exgnn}) \}$ such that
$\dim \Hbg_{\exgnn} = \dmn^{\exgenm}$ for some integer $\exgenm$ for every
$\exgnn$. We can define the achievability with this restriction on codes,
and provided the employed fidelity is $\Fen$ [$F_{\rm p}$], 
we denote the corresponding capacity by $\QCed$ [$\QCpd$]. 

Now we will check the equalities among the four quantities.
Put $\QCe=\Capa$ for accordance with the other three.
It is known\myspcite{barnum00} that
$1-\Fen(\mms{\Hbg},\Enr) \le (3/2)[1-F_{\rm p}(\Hbg,\Enr)]$
for any TPCP map $\Enr$. Hence, $\QCp \le \QCe $,
which is shorthand for `$\QCp(\cA) \le \QCe(\cA)$ for any memoryless
channel $\cA$'. 
From this fact and by definitions, we have
\[
\QCpd \le \QCp \le \QCe, 
\]
and
\[
\QCpd \le \QCed \le \QCe.
\]
Then, all we have to show is $\QCpd \ge \QCe$. This follows from
that the entanglement fidelity $\Fen(\mms{\Hbg},\Enr)$ 
is not larger than the pure-state
fidelity
$(\dim \Hbg)^{-1}
\sum_{\phi \in \cS} \bra{\phi} \Enr(\ket{\phi}\bra{\phi})\ket{\phi}$
averaged over $\cS$,
where $\cS$ is an arbitrary orthonormal basis of $\Hbg$\myspcite{schumacher96}.
In fact, we can reduce $\Hbg$ to a good subspace $\Hbg'\subset\Hbg$ of dimension $\lfloor \dmn^{-1} 
\dim \Hbg \rfloor$ only with negligible loss of the fidelity 
as in the proof of Lemma~1 of \mycite{hamada01g} or as
in Section~V-A of \mycite{barnum00}.
Specifically, 
$F_{\rm p}(\Hbg_{\exgnn}',\Enr_{\exgnn}) \to 1$
for a good choice of $\Hbg'_{\exgnn}\subset\Hbg_{\exgnn}$
provided $\Fen(\mms{\Hbg_{\exgnn}},\Enr_{\exgnn}) \to 1$ as $\exgnn\to\infty$,
where $\Enr_{\exgnn}=\Dbg_{\exgnn}\cA^{\tnsr \varm}\Ebg_{\exgnn}$.
This implies $\QCpd \ge \QCe$, completing the proof.


\end{document}